\documentclass[twocolumn,showpacs,preprintnumbers,amsmath,amssymb,prl]{revtex4}

\usepackage{graphicx}
\usepackage{dcolumn}
\usepackage{bm}

\begin{document}

\preprint{APS/123-QED}

\title{Near-sighted superlens}%

\author{Viktor A. Podolskiy}
\email{vpodolsk@princeton.edu}
\homepage{http://corall.ee.princeton.edu/vpodolsk}
\author{Evgenii E. Narimanov}
\affiliation{Electrical Engineering Department, Princeton University, Princeton, NJ 08544}

\date{\today}

\begin{abstract}
We consider the problem of subwavelength imaging via a slab of a left handed media (LHM) in the presence of material losses. We derive the expression for the resolution limit of LHM-based lens and demonstrate that the area of its subwavelength performance is usually limited to the near-field zone. 
\end{abstract}

\maketitle

The materials with simultaneously negative dielectric permittivity and magnetic permeability, also known as left-handed media (LHM) \cite{veselago}, have recently attracted unprecedented attention \cite{pendry,smith,optExpFocus,shelbyScience,pendryNature,Parazzoli,smithTHz} due to their surprising and often counterintuitive electromagnetic properties. These include reversed Snell's law, Doppler Effect, Cherenkov radiation \cite{veselago}, and unusual nonlinearities \cite{kivshar}. One of the main applications of LHM is the so-called superlens, when a parallel slab of the material with $\epsilon=\mu=-1$ serves as a ``focus-free'' lens {\it with perfect resolution in the far field} \cite{pendry}. Although the implementation of such a device may potentially lead to a tremendous advance in imaging, sensing, fabrication, communications, and related areas, the underlying physics has initiated a lot of controversy\cite{vesperinas,pendryComment,vesperinasAnswer}. 

In the present letter we show that the resolution of the LHM-based lens is strongly suppressed even for a small absorption, inevitable in all modern resonance-based LHM designs \cite{smith,JNOPM2002,podolskiyOptExp,shelbyScience}. We demonstrate that although the near-field performance of the ``super-lens'' is encouraging, the far-field subwavelength resolution is practically unachievable using current techniques. We develop the analytical theory connecting the focal distance of LHM-based lens and its resolution, and solve the above-mentioned controversy. 

The fundamental difference between the conventional and LHM-based imaging is clearly seen when the motion of a wavepacket through the imaging system is considered {\it in the wavevector space}. As a light pulse, represented by a series of different plane waves with the same frequency $\omega$, but different components of the wave vector $k_z$ and $k_x$ (related through a dispersion relation $k_z^2+k_x^2=\frac{\omega^2}{c^2}=k^2$), propagates away from the source, it is subjected to phase and magnitude distortions. The former arise from the phase difference between components with $|k_x|<k$ propagating in different directions; the latter correspond to an exponential decay of so-called {\it evanescent components} with $|k_x|>k$, carrying the information about subwavelength features of the source. Conventional optics is only able to correct the ``phase'' distortions, while the restoration of (already lost) evanescent part of the spectrum is beyond its capabilities. On the contrary, the ideal LHM-based lens represented by a {\it parallel slab} of a material with both dielectric permittivity and magnetic permeability equal {\it exactly} to $-1$, can in principle not only compensate for the phase difference, but also amplify the evanescent fields, leading to a complete restoration of the image \cite{pendry}.

\begin{figure}
\centerline{\includegraphics[width=6cm]{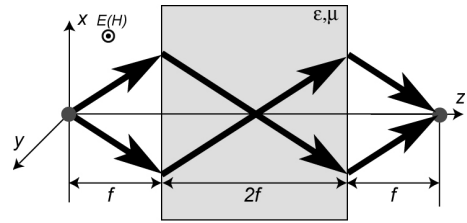}}
\caption{\label{figConfig} LHM-based imaging geometry. The source of the radiation is positioned at the origin of the coordinate system. The arrows correspond to object-to-image ray paths. }
\end{figure}

For simplicity, in this letter we restrict our analysis to imaging of a point slot by a monochromatic wave through the slab of transparent (left-handed) media. We select the $y$ axis of our coordinate system to coincide with the source, and the $z$-axis in the direction of wave propagation. The slab of LHM is positioned between the points $z=f$ and $z=3f$ parallel to the $xy$ plane. The imaging wave has frequency $\omega$ and is polarized with either $E||y$ or $H||y$ (see Fig.~\ref{figConfig}). The generalization of our approach to different geometries is straightforward. 

\begin{figure}
\centerline{\includegraphics[width=8cm]{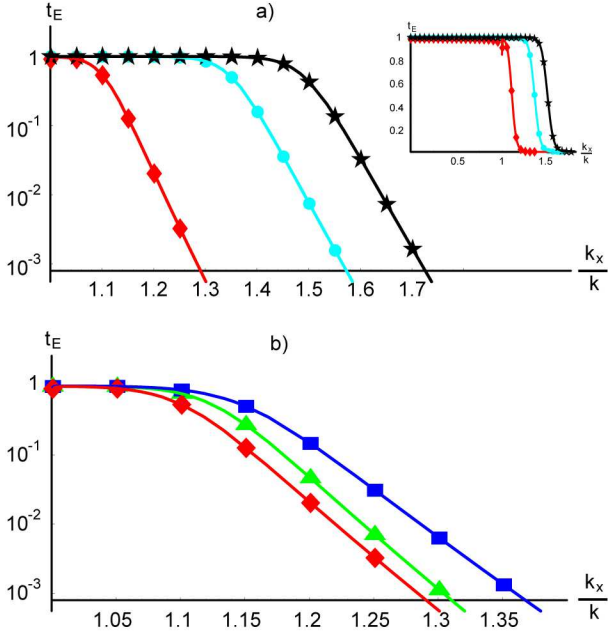}}
\caption{\label{figComp} ``Focus-to-focus'' transmission coefficient of LHM-based lens for $E||y$, $f=\lambda$ as a function of $k_x$ and absorption. Symbols correspond to exact result [Eq.~(\ref{eqTransExact})] (real part of transmission shown), lines represent the low-loss approximation [Eq.~(\ref{eqTrans})];
(a) 
$\epsilon=\mu=-1+10^{-3}i$ (diamond,red), 
$\epsilon=\mu=-1+10^{-5}i$ (circle,cyan);
$\epsilon=\mu=-1+10^{-6}i$ (star,black). 
Inset illustrates the near-rectangular shape of the transmitted spectrum of a point source due to exponential suppression of high-$|k_x|$ (evanescent) spectral components.
(b) The suppression of evanescent components due to ``pure electric'' losses is usually comparable to the one due to ``pure magnetic'' absorption. 
$\epsilon=\mu=-1+10^{-3}i$ (diamond, red), 
$\epsilon=-1$, $\mu=-1+10^{-3}i$ (triangle, green), 
$\epsilon=-1+10^{-3}i$, $\mu=-1$ (square, blue) 
}
\end{figure}

Transmission of an individual (plane wave) component of a light pulse by a parallel slab can be expressed using its dielectric permittivity $\epsilon$ and magnetic permeability $\mu$ via a direct solution of Maxwell equations with corresponding boundary conditions. For our geometry (see Fig.~\ref{figConfig}) we obtain the following transmission coefficients $t_E$ and $t_H$ for the waves polarized $E||y$ and $H||y$ respectively:
\begin{eqnarray}
\label{eqTransExact}
t_{\{E,H\} }= \frac{ e^{i k_x x+i k_z(z-2f)}}{\frac{\kappa_z^2+\{\mu,\epsilon\}^2 k_z^2}{2 i \{\mu,\epsilon\} k_z \kappa_z}\sin(2 f\kappa_z)+ \cos(2 f\kappa_z)} 
\end{eqnarray}
where $k_z=\sqrt{k^2-k_x^2}$, and $\kappa_z=\sqrt{k^2-\epsilon\mu k_x^2}$

\begin{figure}
\centerline{\includegraphics[width=8cm]{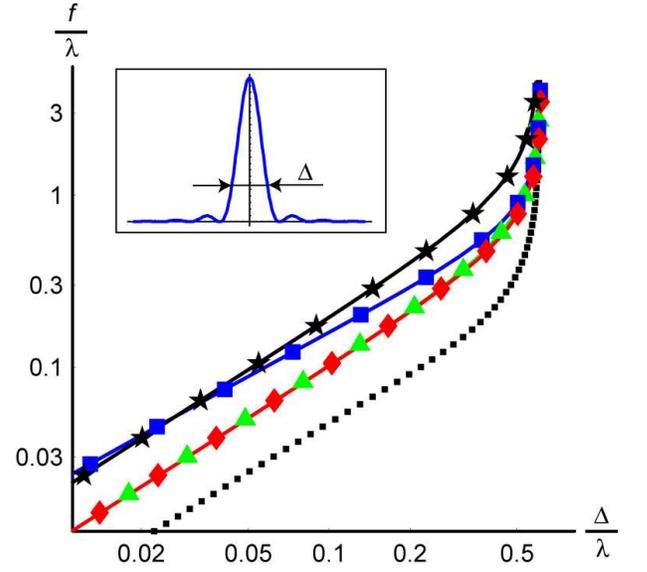}}
\caption{\label{figRes} The maximum focus distance of ``super-lens'' as a function of a desired resolution for the case of imaging of a point slot with electric field $E||y$ (see Fig.~\ref{figConfig}); 
$\epsilon=\mu=-1+10^{-3}i$ (diamond, red), 
$\epsilon=-1$, $\mu=-1+10^{-3}i$ (triangle, green), 
$\epsilon=-1+10^{-3}i$, $\mu=-1$ (square, blue),
$\epsilon=\mu=-1+10^{-6}i$ (star,black); symbols represent the results of exact numerical simulations, lines correspond to Eq.~(\ref{eqRes}). Black points correspond to near-field imaging with conventional lens (reported for comparison). Inset shows the calculated intensity distribution at the focal point along $x$ coordinate, used to determine the resolution $\Delta$. 
}
\end{figure}

As it can be explicitly verified, in the ideal case $\epsilon=\mu=-1$ the magnitude of each wave component at $z=4f$ coincides with its magnitude at the origin ($z=0$) independently of $k_x$. Thus, in agreement with the original proposal by J.~Pendry\cite{pendry} the LHM-based lens may in principle lead to {\it the recovery of the evanescent components} and restore the complete information about the object, in contrast to conventional (phase) optics. It has been recently shown that in lossless media any deviation of $\epsilon$ from $-1$ strongly suppresses the super-resolution \cite{merlin}. However, as we stress here, the lossless media itself represents the ideal case, which can never be realized, especially in the modern resonant-based LHM designs \cite{smith,podolskiyOptExp,shelbyScience}. The presence of any non-zero absorption (described by positive imaginary parts of dielectric permittivity $\epsilon''$ and magnetic permeability $\mu''$) leads to a strong suppression of transmitted waves with large $|k_x|$ (see Fig.~\ref{figComp}). In the limit of small absorption ($\epsilon'', \mu'' \ll 1$), this suppression is exponential and can be adequately described by the following relation:
\begin{eqnarray}
\label{eqTrans}
t_{\{E,H\}}= \frac{1}
{1+\exp({4 f k \chi})
\left[\frac{\{\epsilon,\mu\}''+(1+2 \chi^2)\{\mu,\epsilon\}''}{4 \chi^2}\right]^2},
\end{eqnarray}
where $\chi=\sqrt{\frac{k_x^2}{k^2}-1}$. The excellent accuracy of this expression is illustrated in Fig.~\ref{figComp}, where we compare the transmission coefficient calculated using Eqs.~(\ref{eqTransExact}) and (\ref{eqTrans}) for different losses. 

The suppression of the high-wavevector components limits the resolution of the LHM imaging system since the spatial size of the image $\Delta$ and the spectral width of the corresponding wave packet $\delta$ \cite{resolut} are related through the ``uncertainty principle''
\begin{equation}
\label{eqUnc}
\Delta \cdot \delta = 4\pi\xi, 
\end{equation}
where $\xi$ is a geometrical factor which (although depends on the image and spectrum shapes) approaches the universal limit when the object size becomes much smaller than the corresponding wavelength. The transmitted spectrum in this case has almost-rectangular shape due to exponential suppression of high-$k_x$ components (Fig.~\ref{figComp}~(a), inset). Correspondingly, the geometric factor is obtained from the relation $\rm{sinc}(\pi\xi)=1/2$, which yields $\xi\approx 0.6$. 

The strong dependence of the wavevector cut-off (and the resulting resolution) on the ``focal distance'' $f$ [see Eq.~(\ref{eqTrans})] effectively introduces {\it the maximal focal distance} allowed to achieve a specific detail level via LHM-based lens. Since, as it is shown below, such maximal separation between the {\it subwavelength} object and LHM-lens is typically smaller than the wavelength, the resolution limit of LHM-based optics is somewhat similar to the near-field resolution limit of ``phase'' optics. 

Specifically, substituting the cut-off values of $t_{\{E,H\}}=1/2$ and $k_x=\delta/2=2\pi\xi/\Delta$ [see Eq.~(\ref{eqUnc}), inset in Fig.~\ref{figComp}(a), and Ref.~\onlinecite{resolut}] in Eq.~(\ref{eqTrans}), we obtain the following relation between the focal distance of LHM-based lens $f$ and its spatial resolution $\Delta$: 
\begin{eqnarray}
\label{eqRes}
f_{\{E,H\}}=\Delta\frac{\ln\left[\frac{4\left(\xi^2\frac{\lambda^2}{\Delta^2}-1\right)}
{\{\epsilon,\mu\}''+\left(2\xi^2\frac{\lambda^2}{\Delta^2}-1\right)
\{\mu,\epsilon\}''}\right]}{4\pi\sqrt{\xi^2-\frac{\Delta^2}{\lambda^2}}}
\end{eqnarray}

In Fig.~\ref{figRes} we compare the resolution obtained from Eq.~(\ref{eqRes}) to the one obtained using exact numerical simulations of LHM-lens imaging of a point slot source. We demonstrate that our analytical result is in excellent agreement with exact computer simulations. Eq.~(\ref{eqRes}) is also consistent with the analysis of pure magnetic losses \cite{smithRESOLUT} (note that the actual resolution limit asymmetrically depends on electric and magnetic parts of absorption), numerous numerical simulations of LHM imaging \cite{kivshar04,lagarkov04,ziolkovski,atwater04}, and some recent near-field experiments \cite{sridhar,lagarkov04,eleftheriades04}. 

We also compare LHM-lens resolution with numerically simulated resolution of an ideal ``phase'' optical system. It is clearly seen that while {\it in the near field} LHM-based lens may outperform its ``phase'' analog (see Fig.\ref{figRes}), the logarithmic dependence of $f$ on losses suggests that it is practically impossible to fabricate the long-awaited ``super-lens'' with deep-subwavelength resolution and large (as compared to wavelength) focus distance. 

The developed approach is easily generalized for the case of imaging with cylindrical \cite{pendryCyl} and other types of left-handed lenses, yielding qualitatively similar limiting expressions. 

This work was partially supported by NSF grant DMR-0134736


\begin{thebibliography}{}
\bibitem{resolut}Here the image and spectral width are calculated at ``half-amplitude'' level.
\end{thebibliography}
\end{document}